\documentstyle[epsf,11pt]{article}

\newcommand{\AmS}{{\protect\the\textfont2
  A\kern-.1667em\lower.5ex\hbox{M}\kern-.125emS}}

\title{Time evolution for quantum systems at finite temperature}

\author{E. Mendel and M. Nest\\
  FB Physik, Carl von Ossietzky Universit\"at Oldenburg, \\ 
    26111 Oldenburg, Germany}
\date{}

\begin{document}

\maketitle

\begin{abstract}
  This paper investigates a new formalism  to describe real
time evolution of quantum systems at finite temperature. 
A time correlation function among subsystems will be derived which allows for a
probabilistic interpretation. Our derivation is non-perturbative
and fully quantized. Various numerical methods used to compute the
needed path integrals in complex time were tested and their 
effectiveness was compared.
For checking the formalism we used the harmonic oscillator where
the numerical results could be compared with exact solutions. 
Interesting results were also obtained for a system that presents 
tunneling. A ring of coupled oscillators was treated in order to 
try to check selfconsistency in the thermodynamic limit.
The short time distribution 
seems to propagate  causally in the relativistic case.
Our formalism can  be extended easily to field theories where it remains
to be seen if relevant models will be computable.
\end{abstract}

\newpage
\section{Introduction}

\hspace{5mm}
     Our aim in this work is to develop a non-perturbative formalism that
combines real time and finite temperature for a full quantized theory.    
Actually, it makes little physical sense to talk about time evolution
for {\em the whole system at a finite temperature} (as we would like to avoid
using an abstract external heat bath). So it will be important to keep
in mind that we want to study time-correlations  among small subsystems
of the full quantum system, of which we only know that the large system is in a {\em thermal
state}. The formalism will show its consistency if these correlations
tend to a fixed limit, in the thermodynamic limit of large full systems.

  The original motivation was the understanding of real time processes
at intermediate $T$'s due to instanton tunneling \cite{Smit,Kras}, important
within some field theories to obtain baryogenesis or quark deconfinement.
 The scope has been enlarged as we realize
that one can pose similar questions in many areas of Physics or Chemistry,
where one would like to get time correlations of some degrees of freedom
for quantum systems at a given temperature. For example, the tunneling rate
in ammonia molecules at finite $T$, or spin $t$-correlations in Ising-like 
solids.

  We will assume that the quantum system is in a {\em thermal state} 
described by a density matrix $\rho$ at time zero. We make then a first
measurement on one (or a few) degrees of freedom, the applied operator
being a projector onto the subspace compatible with the first measurement.
We assume here, as in usual quantum mechanics, that the ``collapse of 
the wave function'' in the measurement process, can be described well enough
by the non-unitary projection operator. 
To effectively reduce the {\em kets} and {\em bras} of $\rho$ to the measured subspace, the projector
has to be applied on both sides. A time $t$ later, we make another measurement
of some observable  through the corresponding projector. We will show
that the  correlation function defined in this way has a nice and proper
transition probability interpretation.        
This probability is  gotten  by multiplying  the
probabilities to be in quantum states (according to $\rho$) by the probability
to go from these states (projected onto some states of the subsystem) to some
states at a later time. 

 This contrasts with the usual way to estimate time correlations by just 
computing thermal expectations of operators at two times, as derived for
example from the linear response theory \cite{Fett}. We will show that those 
expressions differ from ours if $\rho$ doesn't commute with the projector, and
do not have a probability interpretation. In the classical limit the operators
commute and both expressions are equivalent. 

Usually, at finite $T=1/\beta$, one just takes the trace over Euclidean time
$\beta$ to calculate expectation values, which is relatively simple to compute with Feynman
Path Integrals. For time correlations we will 
need  a circuit in the complex time plane of the shape:
$\ (0 +i 0) \rightarrow (t - i \epsilon) \rightarrow  (0 -2 i \epsilon) \rightarrow
 (0 -i \beta) $.
The measured time correlations will be obtained by inserting 
suitable operators at the points $ (0 +i 0)$, $(t -i \epsilon)$ and $(0 -2 i \epsilon)$. 
The Path Integrals needed for this complex $t$ will still be doable and converge
with various methods, as we will show in the paper. 
In Section 2 the formalism will be developed and a representation in terms of
the Path Integral formalism in ,,complex'' time will be given. Section 3
outlines and compares three numerical methods to do this integration. Section
4 treats some solved models like the harmonic oscillator, a double well, 
or the interesting case of a ring of N coupled oscillators. Finally
Section 5 summarizes the results and gives an outlook over projected advancements.

\section{Real time formalism}
\hspace{5mm}
  The  question we want to ask is: what is the probability for some observed 
degrees of freedom, of a large quantum system at finite temperature, to make a transition 
to some other observed values after a time $t$.  
We will show that this probability is given by a product of three Green's 
functions in the complex $(t,\beta)$ plane, as indicated in Fig.1, where
one integrates over all unobserved variables. The whole expression can be
casted into a Feynman Path Integral with periodic paths on the complex time 
circuit.

  To fix ideas, let us ask for the probability, $P_t^{\beta}(x_1,y_1)$, 
that if the first particle
(of a system of $N$ interacting  quantum particles) is located close to $x_1$
at $t=0$, it will be close to $y_1$ at time $t$,  the system being at 
temperature $T = 1/\beta$. 

  In order to deduce $P_t^{\beta}(x_1,y_1)$, we start from the knowledge that for
$t=0$ the system is in a {\em mixed state}, described by the thermal density matrix:
\begin{equation}
\rho = \exp(-\beta H) = \sum_n  \exp(-\beta E_n) |n \rangle  \langle n| , 
\end{equation}
with the additional condition that the first degree of freedom be 
around $x_1$, which is obtained by applying the projector 
\begin{equation}
P_{x_1}:= \int_{x_1-a}^{x_1+a} \!  dx_1 |x_1 \rangle \langle x_1| \ * \ {\rm I}_{ 2,..,N} 
\end{equation}
on both sides of $\rho$. The ${\rm I}_{2,..,N}$ means the identity operator
for the coordinates 2 to N.
Note that we have to take a small window $[x_1-a, x_1+a]$ around $x_1$ as otherwise we 
would measure the position with infinite precision implying total delocalization 
afterwards (we can see this effect numerically, in that the $P_t^{\beta}(x_1,y_1)$ goes to a 
constant in $y_1$ for very small windows),
with the projector  still satisfying $P_{x_1}^2 = P_{x_1}$.
 So, for $t=0$ we describe the system by
\begin{equation}
P_{x_1} \: \rho \:  P_{x_1}.
\end{equation}
This operator evolves in time as usual, 
\begin{equation}
 U^+_t \:  ( P_{x_1} \: \rho \: P_{x_1} ) \:  U_t \; \; \; \;  \mbox{   with} \; \;  U_t = \exp(-i H t).
\end{equation}
We measure again at  time $t$, with a position $y_1$ for the first 
particle.  The probability is then:  
\begin{equation}
  P_t^{\beta} (x_1,y_1) = \frac{1}{\rm Norm.} \: Tr \left\{ P_{y_1} U^+_t (
    P_{x_1} \rho P_{x_1} ) U_t \right\}
\label{corr}
\end{equation}
where we have discarded one of the  $P_{y_1}$  due to the cyclicity of 
the trace. This equation has the right probabilistic interpretation as we will see.
The normalization can be taken so that the integration over $y_1$ in Eq.(3) is one,
corresponding to the assumption that the particle {\em is} in the window 
$[x_1-a, x_1+a]$,
or with $Z$, in which case we include the a priori probability to be in that window.
In both cases the integrated $P_t^{\beta}$ is $t$-independent.  

Using energy eigenfunctions  to perform the trace in Eq. (5), one obtains
an expression which allows a simple probabilistic interpretation:

\begin{eqnarray}
P_t^{\beta}(x_1,y_1) \!  \!  \!  &=& \! \!   \int_{x_1-a}^{x_1+a} \!  \!  dx_1  dx'_1 \! 
\int\limits_{i=2,..,N} \!  dy_i dx_i dx'_i \:  \sum_n \exp(-\beta E_n) \:
\langle n | x_i  \rangle
\langle  x_i |  U_t | y_i \rangle
\langle y_i |  U^{\dagger}_t | x'_i \rangle
\langle  x'_i | n  \rangle  \nonumber \\
   &=& \! \!   \int\limits_{i=2,..,N} \! dy_i \: \sum_n \exp(-\beta E_n)
\left| \int_{x_1-a}^{x_1+a} \!  \!  dx_1   \! \int\limits_{i=2,..,N} \!  dx_i 
 \; G_t(y_1,y_i;x_i,x_i) \; \psi_n 
(x_1,x_i) \right|^2
\label{proba}
\end{eqnarray}
This means (as mentioned above) that the {\em probability} to be on an energy level
becomes multiplied by
the {\em probability} to move from this level at the  $\vec{x}$'s, to the 
position states at $\vec{y}$. Note that all the amplitudes from within the window $x_1$
and  the other $x_i$ can interfere with each other as allowed in Quantum Mechanics, as
we have not measured them more precisely. In contrast, the final $\vec{y}$'s
could all have been precisely measured, but we chose not to measure them in Statistical 
Mechanics, so they all add as probabilities.

  It should be mentioned that the correlation function
used frequently in linear response theory (see ref. \cite{Smit}),
\begin{equation}
\langle Q(t)Q(0) \rangle _{\beta} := Tr \{ \exp(-\beta H) Q(t)Q(0) \} ,
\label{usu}
\end{equation}
cannot be interpreted in this way, i.e. as a transition probability.
 Notice that if these $Q's$ were
projection operators Eq. \ref{usu} and Eq. \ref{corr} would coincide in the
classical limit, where $P_{x_1}$ and $\rho$ commute. But in the
quantum case if one  neglects one of the projectors, there is a mixing of amplitudes 
from terms coming from different energy states in the thermal mixture, 
which is clearly incorrect. In fact Eq. \ref{usu} does not even give a real positive number. If
one uses $|Q(t) - Q(0)|^2$ instead, which is positive per construction, 
it still doesn't have a probability interpretation. It should be mentioned that even if one
starts from Eq. \ref{usu}, one needs similar complex t-plane circuits to Fig. 1, as shown
by Semenoff, Niemi, Weiss, Kobes and others \cite{Seme}, when deducing the relevant  
Feynman perturbation rules.

\begin{figure}[hbt]
\setlength{\unitlength}{1cm}
\begin{minipage}[t]{7cm}
\epsfxsize=80mm
\epsffile{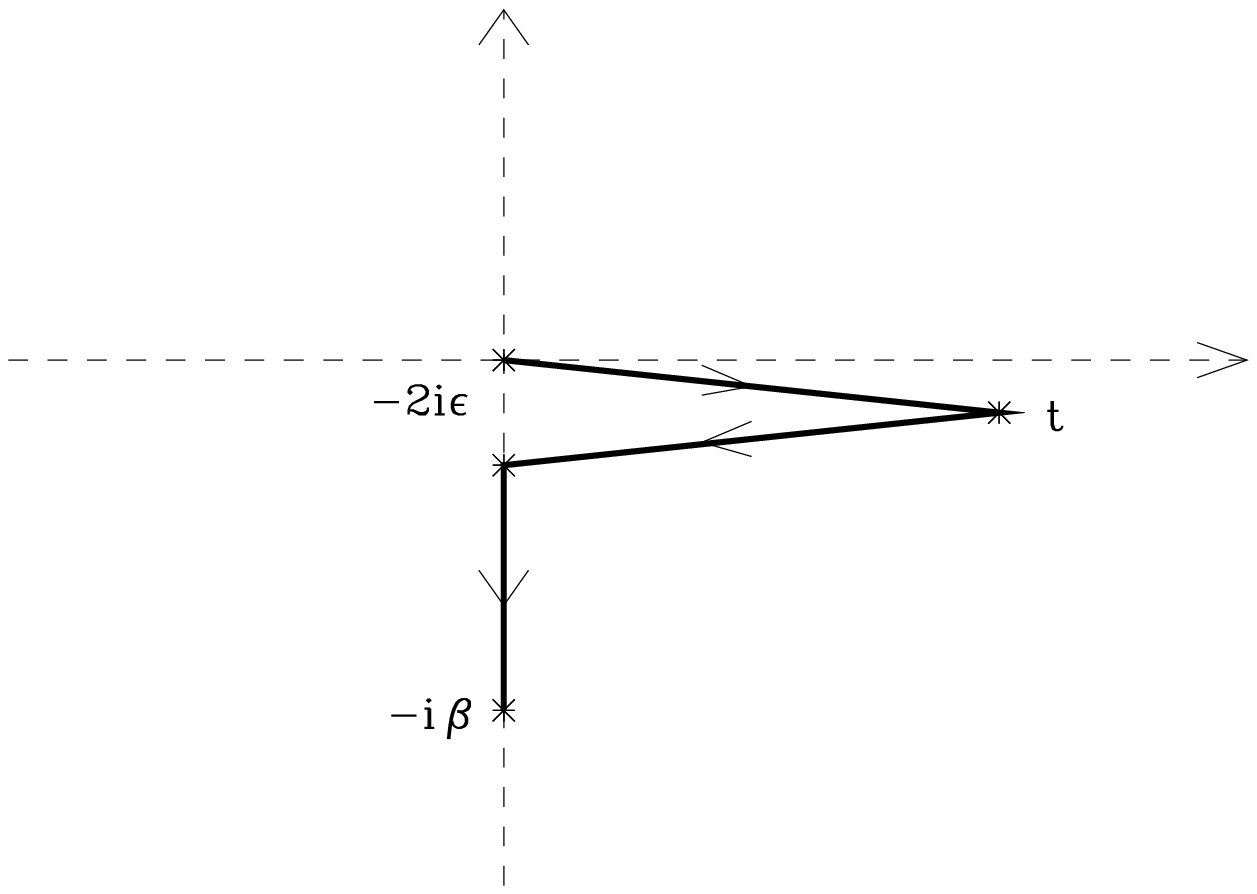}
\par
\caption{Propagation in complex time}
\end{minipage}\hfill
\begin{minipage}[t]{7cm}
\epsfxsize=80mm
\epsffile{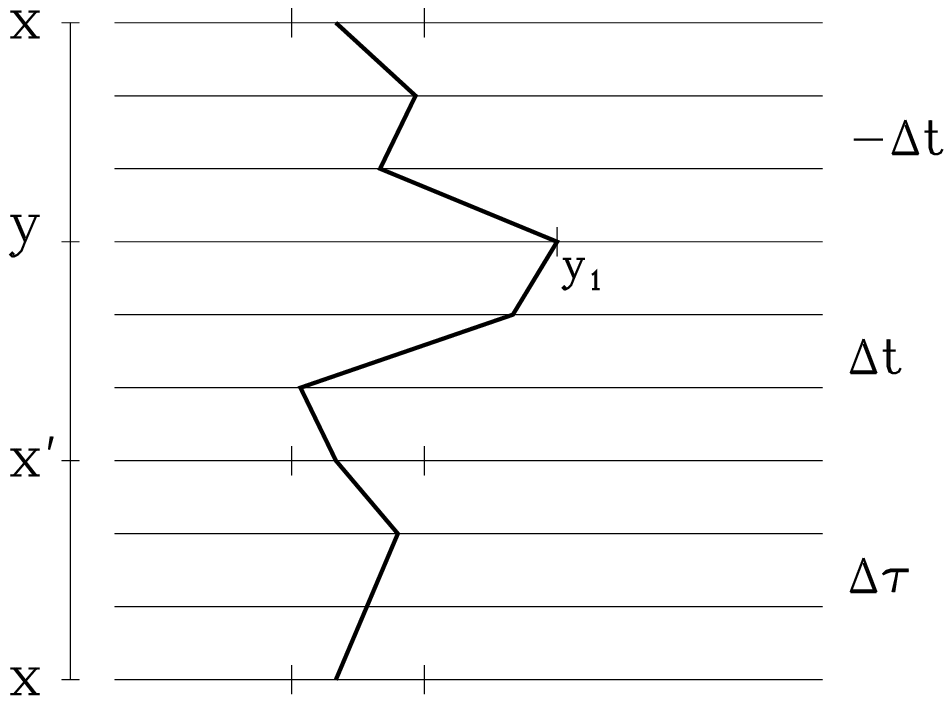}
\caption{one possible path in complex time}
\end{minipage}
\end{figure}

 The evaluation of Eq.(5) can also be done by using position eigenfunctions,
in which case we obtain a product of three Green's functions ($G_\beta \:
G_t^{\dagger} \: G_t$): 
\begin{equation}
P_t^{\beta}(x_1,y_1) =  \int_{x_1-a}^{x_1+a} \!  \! \!  dx_1  dx'_1 \! \!  \!   
\int\limits_{i=2,..,N} \!  \!  dx_i dx'_i dy_i \:  \langle x_1,x_i|\rho |x'_1,x'_i \rangle
\langle x'_1,x'_i|U^{\dagger}_t|y_1,y_i \rangle \langle y_1,y_i|U_t |x_1,x_i \rangle
\label{ggg}
\end{equation} 
This expression is in general easier to evaluate
than Eq. \ref{proba}, as it includes at once all of the energy eigenfunctions. 
This product of three Green's functions represents a
propagation in "complex" time, as shown in fig. 1 (The $\epsilon$ will be
introduced in  Sec. 3.2 , here it is set to $0_+$). Each Green function can
be calculated by a Feynman Path Integral, but the product of the three can also
be expressed by one Path Integral with constrained paths at the windows for the appropriate times as
seen in fig. 2 (the paths are unconstrained in the other not shown dimensions). 

 The generalization to a field theory is in principle immediate,
in fact it is being worked out for the $1 + 1$ Quantum Ising model \cite{Mend3}, where
one is interested in the spatial and time correlations of spins at several temperatures.

\section{Numerical Methods}
\hspace{5mm}
  When using Eq. \ref{ggg}, one has to evaluate the Green's functions and perform
the integrations. It is well known that the Monte Carlo method can be used for
the calculation of the temperature Green's function (i.e. for imaginary
times). Numerical methods for the calculation of real time Greens functions
are still rare. One method for their calculation,
which one of us (E.M.) developed recently, and two other methods were
implemented and tested. In the following we will present them and discuss
their comparative (dis-)advantages.

We will start always from the real time path integral representation of
the Greens functions:

\begin{equation}
G_t(y,x)=\int {\cal D} x \: {\cal D} p \; e^{ i \int_{0}^{t} dt \left( p \dot{q} - H(p,q) \right) }
\label{pi2}
\end{equation}
For two of the methods we will integrate out the momenta to get the usual:
\begin{equation}
G_t(y,x)=\int {\cal D} x \; e^{ i S [ x(t)] }
\end{equation}
To compute this expression  we have to consider discrete $\Delta t$ and check the convergence
to infinitesimal $dt$. The analytical continuation to Euclidean times can be
easily obtained by taking $\Delta \tau = i \Delta t$ in eq. \ref{pi2}.
Actually we can take $\Delta \tau$ to be a fully complex number and move
diagonally in the complex t-plane. In fact it will be useful for improving
convergence to  take slightly slanted lines ($\Delta t - i \Delta \epsilon$) 
for the real time propagation (where $\epsilon / \Delta \epsilon = N_t$ steps) .

\subsection{Matrix Method}
\hspace{5mm}
  This method was first proposed by Dullweber, Hilf and Mendel  in \cite{Mend}.
They have shown that this method works for both real and imaginary times.  
The basic idea consists in considering the propagator for short
times $\Delta t$ only among discrete points in x-space, 
therefore replacing the path integration by  matrix multiplications with itself.
It was interesting to observe that the method converged, even for real times,
with reasonably fine spatial discretizations ($\Delta x < \Delta t /10$) . 
The short time propagator matrix reads:

\begin{equation}
 K_{ij}(\Delta t) \approx \left( \frac{m}{2 \pi i \Delta t} \right)^{1/2} e^{
  i \left[ \frac{m}{2} \left( \frac{x_{t+\Delta t}^i - x_t^j}{\Delta t}
  \right)^2 - V(\bar{x}) \right] \Delta t }
\end{equation}
Here one only needs the short time Lagrangian instead of the action as a function of the
endpoints. The discretization in space delivers the indexation for the matrix  $K(\Delta t)$. 
In this way we obtain

\begin{equation}
 G_t = \left[ K(\Delta t) \right]^N (\Delta x )^{N-1}.
\end{equation}
Using $n$ successive matrix quadratures we obtain larger times according to 
$t=\Delta t \ast (2^n)$. From our numerical experience this has been the fastest method. 
It is the best one to
study systems at deep temperature or their long time behaviour. On the other
hand these matrices do become very large if the system has too many degrees of
freedom needing large amounts of memory. 

\subsection{Monte Carlo Method}
\hspace{5mm}
The path integration could be replaced by a summation over randomly generated paths,
with a Metropolis weighting for the Euclidean pieces of the action. (It is
possible to choose right from the beginning paths for the product of the three
Greens functions). But this summation will not converge, as the amplitudes for
the real time Greens functions all have the same absolute value and differ only
by their phases. This situation can be improved by introducing a small imaginary part
$-i\epsilon$ into the time propagation, so that the paths get a weight and
can be made to converge (to the right answer). One has to be careful though in taking 
the limit $\epsilon \leadsto 0$ as in this process the Monte Carlo convergence gets
rapidly worse. The effect of
this $\epsilon$ is easily seen to be
\begin{equation}
 \exp \left\{ i S^{\Delta t-i\Delta \epsilon}\right\} =\exp \left\{-\Delta
   \epsilon \: H \right\} \exp \left\{i S^{\Delta t }  \right\}
\end{equation}
where we have neglected terms of $O(\Delta \epsilon^2)$ and where the superscripts
in the actions mean the time intervals over which we integrate.
 This means that paths with very
high energy are exponentially suppressed. Unfortunately the convergence slows
down rapidly in the $\epsilon$ to $0$ limit. It does
not seem to be reasonable to use this method for more than 2 or 3 steps of $dt$, but
low temperatures can be reached.
Nevertheless it is the method of choice if one is limited in some problem by
computer memory, as it needs the least with this method.

\subsection{Fast Fourier Transform Method}
\hspace{5mm}
   Here one starts from the path integration over phase space in eq. \ref{pi2}, as explained in
Ref. \cite{Onof}. 
After discretization of space and time, the Greens function can be written as the succession 
of two  Fourier Transforms per time step (we have generalized their folding with a given wave function):
\begin{eqnarray}
 G_{\Delta t}(x_1,\tilde{x}_0) & = & 
\int dp_1 dx_0 \exp \left\{i \left( p_1 (x_1-x_0) 
-\frac{p_1^2\Delta t}{2m} -V(x_1,x_0) \Delta t \right) \right\} \delta (x_0-\tilde{x}_0) \\
{} & = & {\cal F}^{-1} \left[ \hat{T} {\cal F} \left[ \hat{V} \delta  
(x_0-\tilde{x}_0) \right] \right]
\end{eqnarray}
Here ${\cal F}$ means Fourier Transform and $\hat{T}$ and $\hat{V}$ are the
exponentiated
kinetic/potential energies, respectively. In other words, the propagator for
one time step from a fixed position $\tilde{x}_0$ to some arbitrary point $x_1$ is
achieved by a twofold Fourier transform, each of which can be computed effectively
with the Fast Fourier Transform (FFT) method. Iterative application of this
operator ${\cal F}^{-1} \hat{T} {\cal F} \hat{V}$ yields the propagator
for more time steps (only one $\Delta t$ gain in one iteration in contrast to the matrix method).

For general potentials $V(x_1,x_0)$, where the "mid point rule" is hard to disentangle (for the FFT), it is
better to use the alternative discretization:

\begin{equation}
 \exp\left( V(x_1,x_0) \right) \leadsto \exp\left( V(x_1)/2 \right) \exp
 \left( V(x_0)/2 \right)
\end{equation} 
where the term in $x_1$ can be extracted to the front of the fourier transforms above.
 
  Here too an $\epsilon$ can improve the results, as the
fourier transform assumes the transformed function to be periodic and an
imaginary $\epsilon$ forces the exponentiated energies to fulfill this requirement.

This method has moderate RAM requirements
as only vectors have to be stored, but the computational time increases
linearly with the number of $\Delta t$'s. In principle this method should be faster
than the matrix method in problems with higher number of degrees of freedom, but
in practice this remains to be seen as the FFT has to be done for each of the
"spectator" coordinate anew and the overhead for each FFT compared with a matrix
multiplication is considerable. 

\section{Solved models}

\subsection{Harmonic Oscillator}
\begin{figure}[hbt]
\setlength{\unitlength}{1cm}
\begin{minipage}[t]{7.5cm}
\epsfxsize=80mm
\epsffile{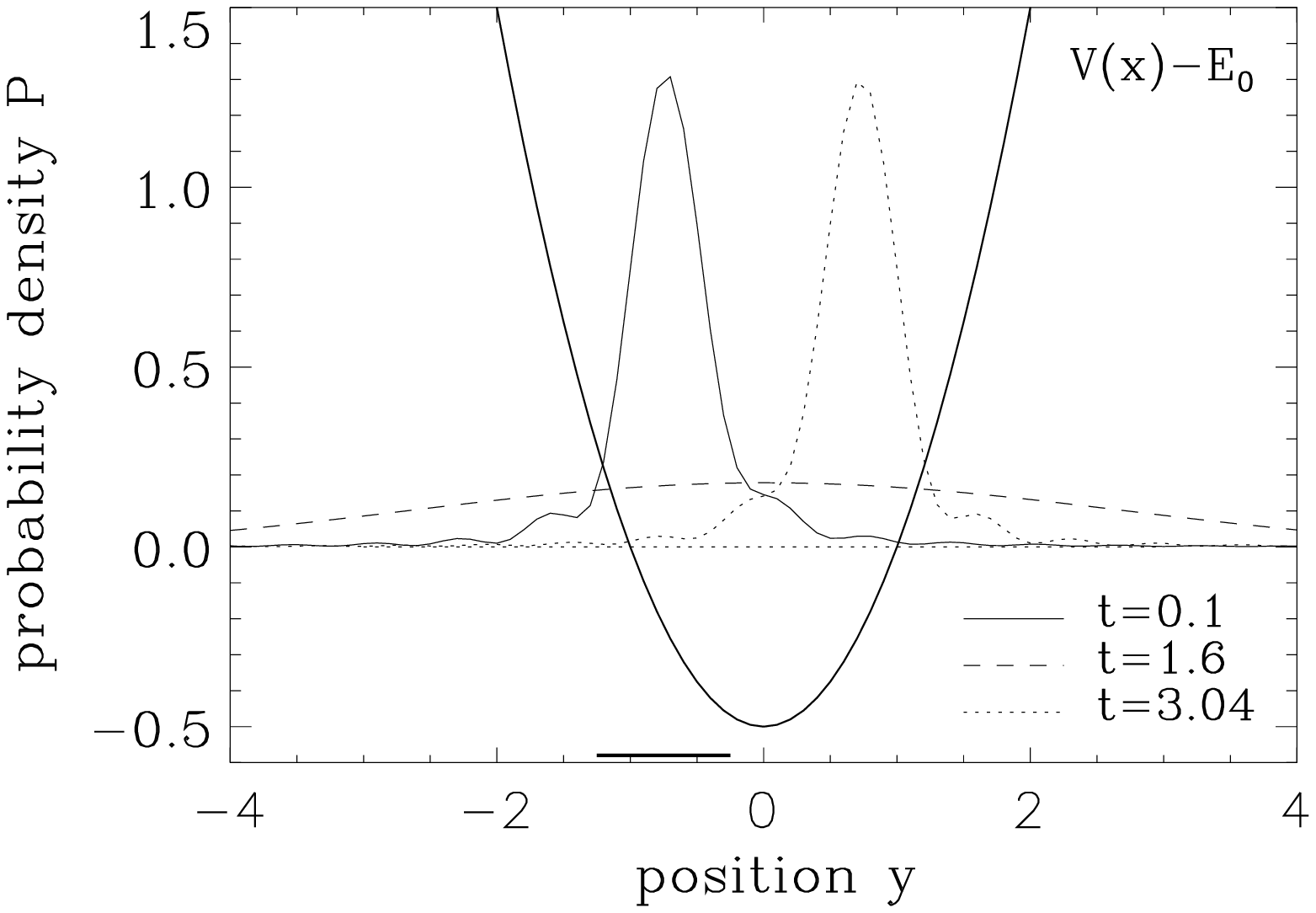}
\par
\caption{Probability  to be at $y$ for an h.o. for times $t \approx$ 
$0$, $\pi/2$ and $\pi$, at $T=1$. Exact Green's functions were used.}
\label{ho1} 
\end{minipage}\hfill
\begin{minipage}[t]{7.5cm}
\epsfxsize=80mm
\epsffile{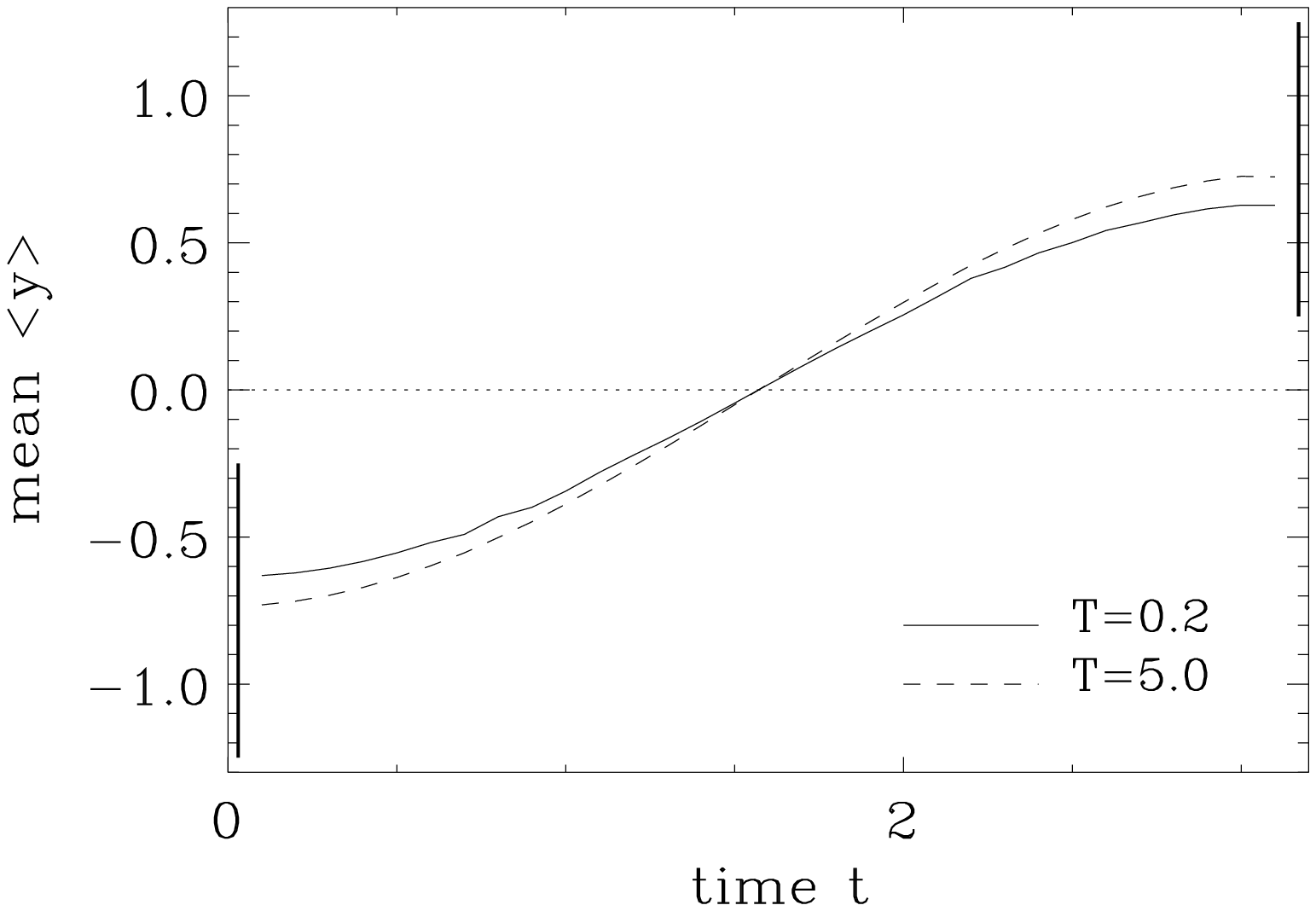}
\par
\caption{Mean of the wave-packet as a function of time, two
  temperatures. Vertical bars are the initial (mirrored) windows.}
\label{ho2}
\end{minipage}
\end{figure}

\hspace{5mm}
   We applied these methods  to compute time evolution in quantum systems at
finite temperature to a variety of systems. We will first treat the harmonic 
oscillator as the evolution of the probability distribution  can be 
compared in this case with analytical results. 
Let us assume that a particle in a harmonic oscillator potential
(with $\omega=1, m=1$) is for $t=0$ in
the interval $[-1.25;-0.25]$ (this starting interval is indicated by a horizontal line in fig. \ref{ho1}). 
The evolution of the probability density is depicted
at three times, so that one can imagine the wave packet oscillating in the
potential. Note that the wave packet at $t=3.04\approx\pi-0.1$ is (as
expected) symmetric to the distribution at $t=0.1$. The extreme broadening at
the intermediate time is due to the chosen high temperature, so that also
energy eigenfunctions of higher levels contribute to the result.
 If we decrease the temperature, we obtain a more localized probability distribution.
To make sure that our correlation function reproduces
indeed the behaviour of a h.o. the movement of the mean of the
probability distribution was calculated (fig. \ref{ho2}). The result is a cosine,
with somewhat larger oscillation amplitude for a higher temperature, as expected.

   For this model with just one degree of freedom, the coherence of the signal 
is maintained (with contributions from  various states), as in these simple
cases each mode does not get perturbed by others, the  Boltzmann distribution
of states being simply postulated.
 In more complex systems, for which this formalism is really being developed, 
the other degrees of freedom of the system  form the
heat bath for the degree of freedom which is being measured and this simple
coherence will be lost. In the case of many degrees of
freedom, the thermodynamic limit should be consistent with the Boltzmann 
distribution. 

  The Green's functions for the h.o. are explicitly known (see e.g. \cite{sch}), so
one can assess the three numerical methods of Sec. 3. The Monte-Carlo method
seems to be unsuitable for more than 2-3 time-steps, even with a fairly large
$\epsilon$. Apart from that, it converges well for  low temperatures (i.e. large
numbers of $d\tau$) with little RAM requirement. Better results and with 
smaller errors due to the $\epsilon$ convergence factors are obtained  with the other two
methods. For large times the matrix method is the fastest, even though larger amounts of
memory have to be used.

As a last test we proved that the total probability turns out to be
 conserved, to a good approximation for discrete $\Delta x$, as expected.

\subsection{Double Well Potential}
\begin{figure}[hbt]
\setlength{\unitlength}{1cm}
\begin{minipage}[t]{7.5cm}
\epsfxsize=80mm
\epsffile{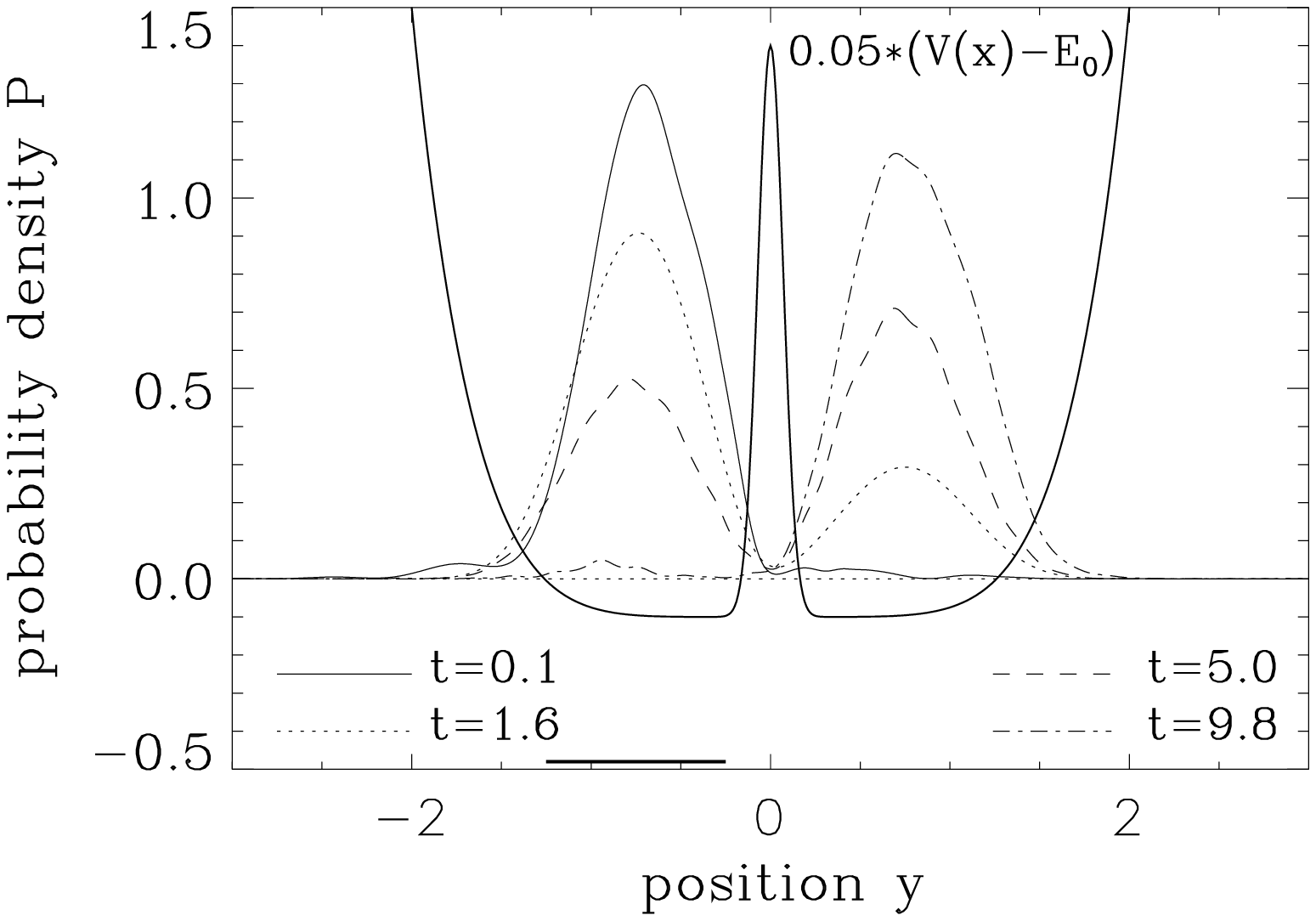}
\par
\caption{Probability  to be at some $y$ for a double well for a low temperature.} 
\label{dw1}
\end{minipage}\hfill
\begin{minipage}[t]{7.5cm}
\epsfxsize=80mm
\epsffile{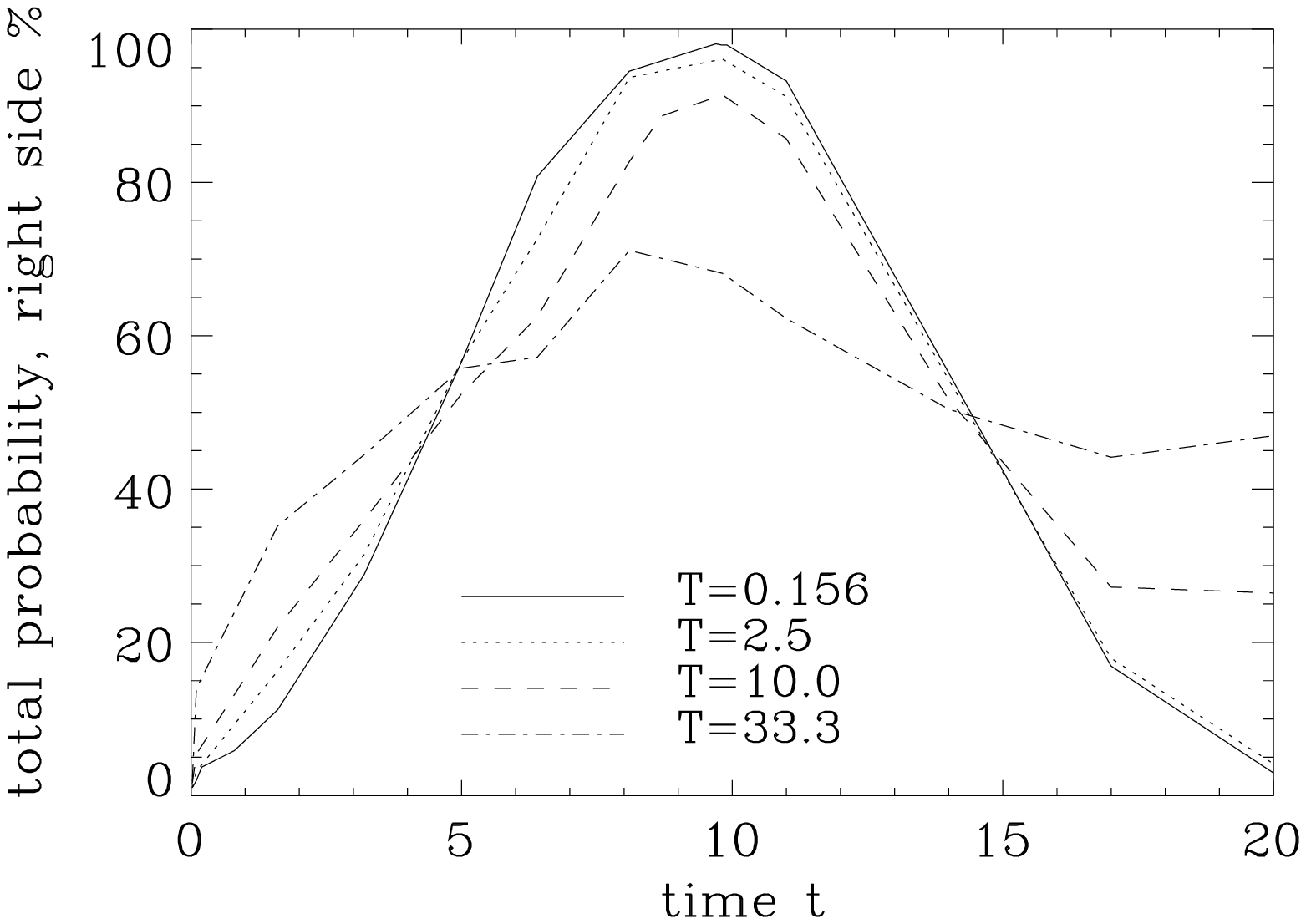}
\par
\caption{Total Probability to be in the right side well as a function of time for
  different temperatures.}
\label{dw2}
\end{minipage}
\end{figure}

\hspace{5mm}
  Here we wanted to study quantum tunneling at finite temperature.  We used a 
nearly box shaped $x^6$-potential with an added Gaussian
barrier (height $E_{max}=30$). Let
us first look at the low T case. At $t=0$ the particle is in a window located on the
left side. Fig. \ref{dw1} shows the temporal evolution. The horizontal bar is
again the initial interval. If only the lowest two energy levels 
contribute reasonably to the distribution, the result is a nearly 100 \%
tunneling after a time $T_{\rm tun} \approx \frac{\pi}{E_1-E_0}$, and
due to the symmetry half of its periodicity. Fig. \ref{dw2}
shows the
total probability to be on the right half as a function of time for several
temperatures. The periodicity of the low-T-case ($T_{\rm tun} \approx 10$) is
apparent. If there are higher wavefunctions that interfere, this property is
destroyed and the total probability after some time to be on each side tends to 1/2. 

While for higher T there is a sizable probability to be on an energy level
above the potential barrier, this is not the case for low T
(e.g. $T \approx 0.156: \; \; p(E>30)<10^{-78}\%$). Thus, semi-classical
perturbation theory as in \cite{Kras}, \cite{grr} cannot be used here. Their
ansatz ({\em for calculating sphaleron-rates in electroweak field theory}) 
was to start with a quantum thermal ensemble of field configurations and then calculate the 
classical time evolution for each of them. Thus only sphaleron-like transitions to other minima
are possible. But for temperatures far below the sphaleron-energy the
behaviour of the system is governed by tunneling processes and therefore the
calculation must be fully "quantum".

\subsection{N coupled oscillators}
\begin{figure}[hbt]
\setlength{\unitlength}{1cm}
\begin{minipage}[t]{7.5cm}
\epsfxsize=80mm
\epsffile{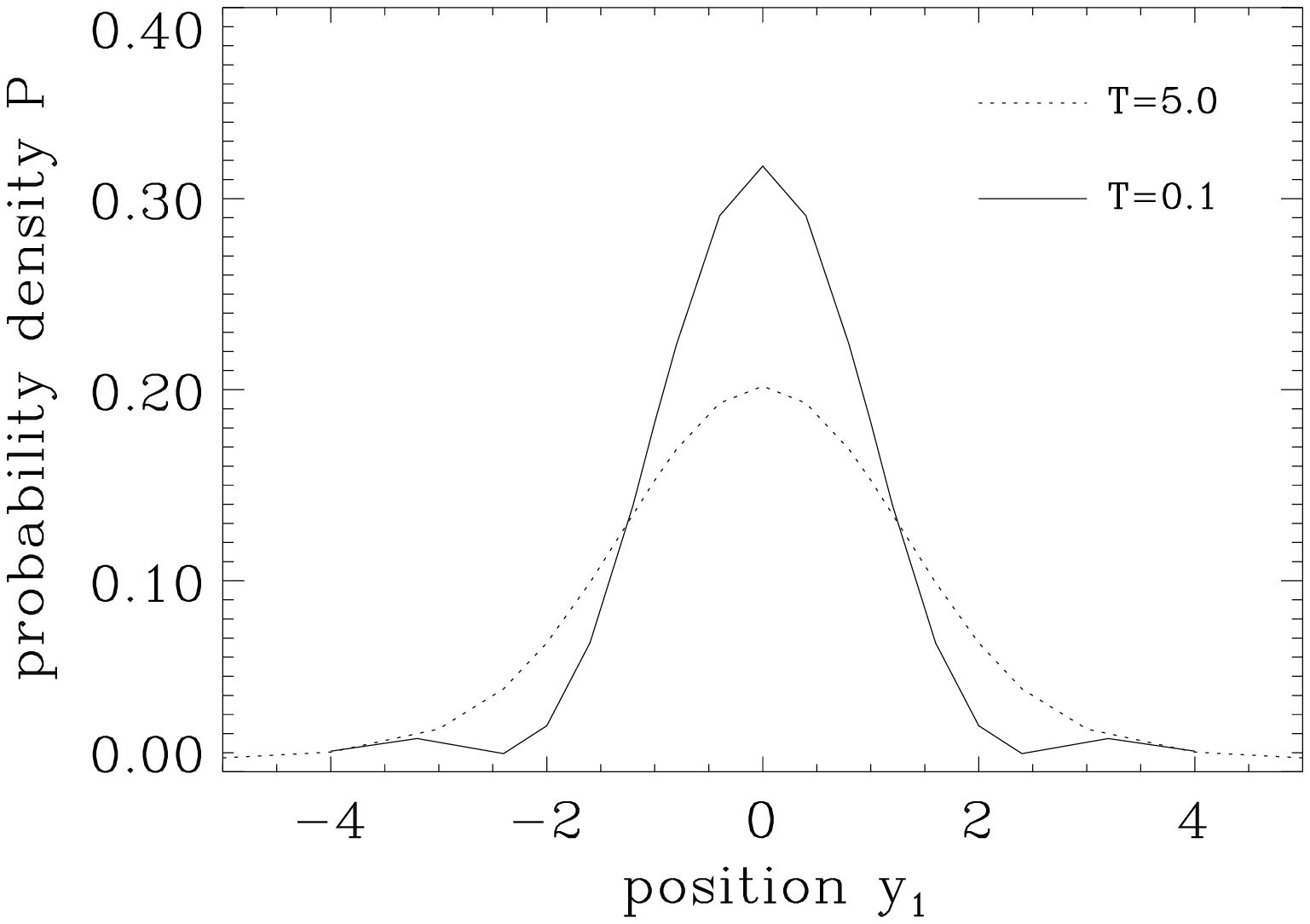}
\par
\caption{Probability  to be at $y$ for particle 1 of a ring of four particles,
    for time $t=0.5$.}
\label{ring1} 
\end{minipage}\hfill
\begin{minipage}[t]{7.5cm}
\epsfxsize=80mm
\epsffile{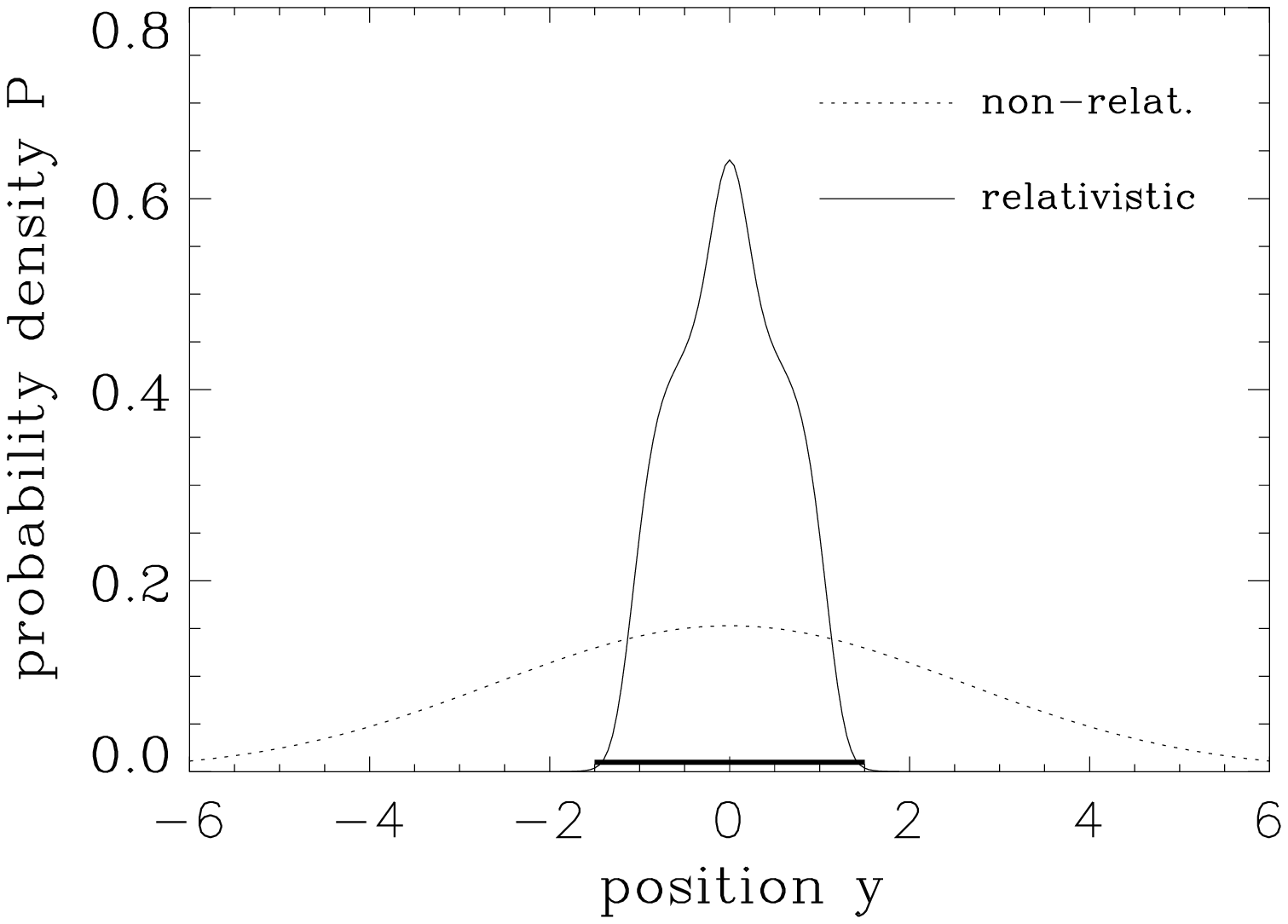}
\caption{Comparison of the relativistic and non-relativistic free
particle cases, with t=1 and T=5.}
\par
\label{relat}
\end{minipage}
\end{figure}

\hspace{5mm}
  The most interesting case that we have solved in this work is the N particle
chain of coupled oscillators on a circle, with transversal oscillations. 
This is the first genuine finite temperature system that we treated in the
sense that the other degrees of freedom form a heat bath for the measured
one. We have been able to solve the Green's function for this model "almost" analytically by going to 
normal modes:

\begin{equation}
 Q_0 = \frac{1}{\sqrt{N}} \sum_{j=1}^{N} q_j
 \label{sp}
\end{equation}

\begin{equation}
 Q_i^{(c)} = \sqrt{\frac{2}{N}} \sum_{j=1}^N q_j \cos \left( \frac{2 \pi i
  j}{N} \right) \hspace{2cm} i=1,\ldots,N/2
\end{equation}

\begin{equation}
 Q_i^{(s)} = \sqrt{\frac{2}{N}} \sum_{j=1}^N q_j \sin \left( \frac{2 \pi i
  j}{N} \right) \hspace{2cm} i=1,\ldots,N/2-1   
\end{equation}
The prefactors are chosen so that the determinant of the coordinate transformation matrix
is one. From now on the two indices of $Q_i^{(c,s)}$ and $P_i^{(c,s)}$
are merged to a single one which runs over all possible cases.
In these new coordinates the Hamiltonian is separable:

\begin{equation}
 H = P_0^2 \; + \;  \sum_i \frac{\omega_i^2}{2} \left( Q_i
  \right)^2 = H_0 \; + \; \sum_i H_i
\end{equation}
where

\begin{equation}
 \omega_i = \sqrt{2-2 \cos \left( \frac{2 \pi i}{N} \right)}
\end{equation}
$H_0$ is the center of mass Hamiltonian, the $H_i$ are related to the phonons.
The Green's functions can then be decomposed:

\begin{eqnarray}
 G_t(Q_0 \ldots Q_{N-1} ; Q_0' \ldots Q_{N-1}')
 & = & \langle Q_0 \ldots Q_{N-1} | e^{-i\hat{H}t} | Q_0' \ldots Q_{N-1}'
 \rangle \nonumber \\
 {} & = & \langle Q_0 | e^{-i\hat{H}_0t} | Q_0' \rangle
 \prod_i \langle Q_i | e^{-i\hat{H}_it} | Q_i' \rangle \nonumber \\
 {} & = & G_t^{{\rm Free}} (Q_0,Q_0') \prod_i G_t^{{\rm HO}}(Q_i,Q_i')
\end{eqnarray}
The corresponding operators satisfy the usual commutation relations:

\begin{equation}
\left[ \hat{Q}_i,\hat{Q}_j \right] = \left[ \hat{P}_i,\hat{P}_j \right] = 0
\end{equation}

\begin{equation}
\left[ \hat{Q}_i,\hat{P}_j \right] = i \delta_{ij}
\end{equation}
Due to the linearity of the transformation and the Jacobian being 1, one
can show,

\begin{equation}
G_t(\vec{q},\vec{q}') = \left. G_t(\vec{Q},\vec{Q}') \right|_{\vec{Q}=
\vec{Q}(\vec{q}) \atop \vec{Q}'=\vec{Q}'(\vec{q}') }
\end{equation}
in words, the Green's function in the original variables (needed in the integrations) is just
given by the solved Green's functions in the new ones, with the new variables taking the
values $\vec{Q}=\vec{Q}(\vec{q})$.
The integration of Eq. (\ref{ggg}) over the old $q$ is over an ,,area'' which is $I \! \!
R^{3N-3}\times I^2$ ($I$ being the initial window). We performed these integrations
numerically and chose $N=4$. Particle number 1 was assumed to be for
$t=0$ in the interval $I=[-0.625;0.625]$. The probability density
$P(y_1,t=0.5)$ was then calculated for two temperatures (see fig. \ref{ring1}). As
expected, the higher temperature leads to a broader distribution, i.e. to a
faster spreadening. Fig. \ref{ring1} needed approximately 12 h of computing time on the
vector processor of the RRZN in Hannover. Thus an increment in the number
of particles would be possible only if one applies approximation techniques or
performs some of the integrations analytically. It would be interesting in the
future to study to which extent we have reached the thermodynamic limit in the
sense that if we have a very large circle of $N$ oscillators the probability
to find the first (or a close neighbour) in some position after some time, 
should not change when we go to $N+1$
oscillators. If this turns out to be the case we will have a good check on our
formalism, as the rest of the degrees of freedom really form the heat bath
and the theory is selfconsistently correct.

\subsection{Relativistic Aspects}
\hspace{5mm}
   As can be extracted from  the previous figures,  the probability density broadens with
speed $v > c$ (we set $c=1$). This is no surprise, as we used
ordinary non-relativistic Quantum Mechanics. To check if our correlation
function can be extended to the relativistic regime, we chose the
Hamiltonian

\begin{equation}
H_{\rm rel} = \sqrt{p^2+m^2} ,
\end{equation}
for a relativistic free particle. We had to use the FFT-method as
only there the Hamiltonian formalism  not yet integrated over momenta (instead of the Lagrangian) can be directly used.
Fig. \ref{relat} compares the result with the free non-relativistic
distribution. The horizontal bar is again the initial window, but it
was switched on with an additional window function, taken as
$\exp(-3x^4)$. The need for this smoothing is due to the well known problem 
of possible pair creation for relativistic fields, if a sharp
rectangular window function is used. A high temperature was used to
stimulate the fast dispersion due to high momenta modes. We
can see that the relativistic wave packet is much more localized. An
exponentially small violation of causality can nevertheless be found
here. The reason for it is that a fourier transform in a compact
interval (the initial window) has necessarily components with negative
frequencies (related to a small probabilty of pair creation), which
cannot be included in our models in first quantization. Accordingly the
small exponentially suppressed violation is entirely present immediately after the first measurement
and does {\em not} increase with time as the probability broadens, as we have carefully checked for our system.
 
\section{Conclusions}
\hspace{5mm}
  We have presented a formalism that describes real time evolution for
some degrees of freedom in quantum systems at finite temperature. The
description is necessarily of probabilistic nature as at finite $T$
we are not in a pure state and one is not allowed to mix coherently 
amplitudes from different states in the thermal mixture. It is therefore
doubtful if the usual time correlation function in thermal states will
be directly checkable experimentally, unless we are close to the 
classical limit where $H$ and $\rho$ commute. 

  The probabilistic formalism for the time evolution can be
expressed by a product of three Greens functions, one in Euclidean
and two in real time, where one integrates over unobserved degrees
of freedom. In the path integral formalism this corresponds to
integrating over periodic paths in a complex time circuit, where
we pinch the paths to the observed values (or ranges) at three (complex) times.

  We presented various calculational methods to compute the required
path integrals for complex time lines and showed that each one has its
advantages, all converging eventually to the same answer. The matrix
method is the fastest one, at least for low dimensional problems, requiring 
the highest memory resources. The FFT method needs less memory but takes
a linear computing time with $t$ or $\beta = 1/T$. The Monte Carlo
method converges very slowly for real time propagation (no problem for large 
$\beta$) , but could
still be used for very short time evolution, with least memory
requirements. For all three methods it was useful to include a small
imaginary $\epsilon$ to the time propagation, which acts as an ultraviolet
regulator.  

  We have first solved models with one degree of freedom in order to
check the computational methods and gain confidence with this formalism,
observing interesting phenomena as temperature dependence in the
spreading probability, tunneling processes or near causality in the
naive relativistic limit. The real consistency proof for our formalism
should come in the thermodynamic limit of many degrees of freedom, in which
we can assume that the initial quantum preparation of a small subset of
the large quantum system will not bring the large system out of thermal equilibrium.
Therefore, in this limit the probability density t-evolution should not
change when we enlarge further the size of the system, the large quantum
system providing the thermal bath for the correlation measurement.
  With the ring of $N$ oscillators model, which we could solve analytically
for the Greens functions, we were able to solve the needed integrations over
the spectator variables only up to 4 oscillators. This gave interesting
results in its own right but we could not yet test the thermodynamic limit.
  Recently, a $1 + 1$ dimensional Ising is being tested with this formalism
\cite{Mend3}, in order to check the thermodynamic limit. In this simplest
field theory one only has two values for the spin at each site, making
the problem computable with the matrix method. The transfer matrix has to
be carefully tuned in its couplings in order to reach a continuum $\Delta t \rightarrow 0$
limit. First results seem to confirm convergence of spin-correlation
functions in time to limiting curves when one enlarges the spatial number of 
spins, thus indicating a thermodynamic limit. It remains to be seen if for
more complicated field theories with interesting instanton effects, one can 
compute the real time pieces at least for short times for this formalism.
  
  There are several aspects of this formalism still under study. For example
the generalization from projector operators as observables, to more general
operators. Also interesting is a better understanding of the linear response
approximation starting from this formalism. Finally, one would like to have
a better hold of the measurement process in quantum theories, to be included
concretely into this formalism to study time evolutions.

\subsection*{Acknowledgments}
 
We thank L. Polley for useful discussions and the RRZN in Hannover for
supercomputer time.


\begin{thebibliography}{00100}
\bibitem{Mend3} E. Mendel, ``Real time correlations at finite Temperature for the Ising model'', in prep.,
 to be presented at the Lattice 98 Conf. in Boulder, Colorado.
\end{thebibliography}
\end{document}